\documentclass[12pt]{article}
\usepackage[utf8]{inputenc}

\title{Renormalizability, fundamentality and a final theory: The role of UV-completion in the search for quantum gravity}
\author{Karen Crowther and Niels Linnemann}
\date{To appear in the British Journal for the Philosophy of Science}

\usepackage{natbib}
\usepackage{amsmath}
\usepackage[margin=1.2in]{geometry}
\usepackage{graphicx}

\usepackage{tikz}
\usetikzlibrary{shapes,arrows}

\begin{document}

\maketitle


\begin{abstract}

    Principles are central to physical reasoning, particularly in the search for a theory of quantum gravity (QG), where novel empirical data is lacking. One principle widely adopted in the search for QG is UV completion: the idea that a theory should (formally) hold up to all possible high energies. We argue---\textit{contra} standard scientific practice---that UV-completion is poorly-motivated as a guiding principle in theory-construction, and cannot be used as a criterion of theory-justification in the search for QG. For this, we explore the reasons for expecting, or desiring, a UV-complete theory, as well as analyse how UV completion is used, and how it should be used, in various specific approaches to QG.  
    \\\\\textbf{Additional keywords:} Spacetime; Renormalization group; Effective field theory; Unification; String theory; Asymptotic safety 
\end{abstract}

\newpage

\tableofcontents

\newpage

\section{Introduction}
The problem of quantum gravity is to find a theory (QG) that describes the phenomena at the intersection of general relativity (GR) and quantum field theory (QFT).\footnote{\label{phen} Most characteristically, GR and quantum theory are both necessary to describe a particle whose Compton wavelength is equal to its Schwarzschild radius---occurring when these quantities are approximately equal to the \textit{Planck length} ($l=l_\text{P}=\sqrt{\hbar G}/{c^3}\approx 1.62\times 10^{-35}\text{m}$), and the particle is of Planck mass ($m=m_\text{P}=\sqrt{\hbar c/G}\approx 1.22\times 10^{19} \text{GeV}$). Other phenomena whose full explanations are expected to be provided by QG include black holes and the cosmological singularity of the big bang.} What else the theory describes is left open, as is the form of the theory. This freedom is a consequence of being relatively unconstrained by empirical data---not only are there no data that QG is definitely required to explain, but none of the approaches appear, at least at present, to be experimentally testable. Instead, the search for QG is guided by \textit{principles}. One of these is the idea that QG be \textit{UV-complete}, meaning that the theory is predictive at all possible high-energy scales (or, equivalently, all short distance scales); i.e., it is a description of the complete ultraviolet (UV) regime. Here, we explore and evaluate the reasons for thinking that QG should be UV-complete, and the roles that UV-completion plays as a principle of theory-construction and theory-justification in particular approaches to QG. Although UV-completion is widely thought to feature centrally as a principle of QG, we find that it does not actually play the important role attributed to it within many particular approaches. Furthermore, we argue that UV-completion is unnecessary and artificial as a guiding principle in modern physics (though it may nevertheless aid the search for QG, by acting as a heuristic motivation and aspiration). We also argue that UV-completion should neither be taken as a postulate of QG, nor a criterion of theory acceptance. In short---\textit{contra} standard belief---QG should not be assumed to be UV-complete.

\paragraph*{}

Before we begin, we make explicit our definitions of \textit{QG} and \textit{UV completion}, and briefly outline two closely-related principles, \textit{finiteness} and \textit{mathematical consistency}. By ``QG'' we mean any theory that satisfies the criteria that are taken to define QG. Of course, however, there is no universally established set of criteria (and even some of the specific research programs lack a definite, well-articulated set of criteria), and so what would constitute QG is an open problem. The aim of this paper is to help with precisely this issue. We argue that the set of criteria has been largely misinterpreted as including the criterion that the theory be UV-complete, when actually the criterion is not only unnecessary, but also, in a sense, unreasonable---unless one includes the additional, non-necessary criterion that the theory be a final, unified ``theory of everything''. The notion of UV completion adopted here is the claim that a theory is \textit{formally}\footnote{This is important, because a theory being UV-complete need not mean that its predictions are \textit{actually} meaningful (i.e., correct, or useful) at all high energy scales.} predictive up to all (possible) high energies. The vast majority of classical theories are UV-complete; QFTs, however, are \textit{not} standardly UV-complete ``out of the box''. As explained below (\S\ref{renuv}), infinities arise in QFTs, and these theories need some mathematical surgery in order to render them predictive \textit{at all}. Some QFTs are then found to be UV-complete, and some are found to not be. Those that are found to \textit{not} be UV-complete are called \textit{effective}---they are (usefully) predictive only at low energy scales compared to some high energy scale ``cutoff'', $\Lambda$. At energies approaching $\Lambda$, the theory ceases to be useful, and eventually breaks down.

\paragraph*{}

An early attempt at quantizing GR puts GR in the framework of perturbative QFT (perturbative quantum GR), in which context it falls into the not-UV-complete class of theories, with the cutoff energy $\Lambda$ equal to the Planck energy (\S\ref{renuv}). This theory is thus considered an effective theory of QG---reproducing the results of QG at low energies. As we explain (\S\ref{Approaches}), many physicists actually consider the Planck-scale breakdown of perturbative quantum GR to be \textit{the} problem of QG! Additionally, it has been interpreted as evidence of the existence of both a fundamental length (i.e., an operational limit on experimental resolution), as well as the breakdown of (continuous) spacetime itself (\S\ref{disc}). It has also been taken as motivation for the requirement that QG be UV-complete (\S\ref{why}). While we acknowledge that this lack of UV completion serves as a significant motivational heuristic in the search for QG, we find that these other claims are not well-founded (\S\ref{why}), and that there are reasons to caution against using UV completion as a guiding principle in this context (\S\ref{whyno}).

\paragraph*{}
The idea of UV completion (as defined) relates to that of \textit{finiteness}: a theory that is finite (i.e., does not contain, nor lead to, divergences)  in its observables  is normally UV-complete. The converse, however, does not have to hold: divergences might pop up in the infrared (IR) as well as in the UV, so UV completeness alone cannot guarantee finiteness.\footnote{Theoretically, however, it is possible that a UV-complete theory might include singularities as a genuine prediction of what nature is like.} That a theory be finite is requisite for its being mathematically well-defined (discussed next). There is also a common perception that a theory must be finite in order to be predictive, but this is incorrect---there are theories that contain divergences yet are predictive within certain regimes (e.g., classical electrodynamics), or can be rendered predictive within certain regimes (e.g., quantum electrodynamics). Nevertheless, one might argue that finiteness, rather than UV completion, is the appropriate criterion to use in the search for QG, or that finiteness is really what is being sought by the appeal to UV completion. The concept of UV completion overlaps completely with that of UV-finiteness, and is thus a precondition for finiteness overall. So, our arguments in this paper, by demonstrating that the (weaker) criterion of UV completion is not necessary for QG, also hold against the (stronger) criterion of finiteness. Just as QG does not necessarily have to hold up to all (possible) high energies when it is really Planck scale physics that is of interest, the theory does not, strictly speaking, have to be finite, either---so long as any divergences do not affect the practical ability to use the theory in the required regime. If finiteness is sought in order to ensure predictiveness, then it is not only unnecesssary, but ill-motivated. This is also the case is finiteness is sought in order to ensure \textit{mathematical consistency}.

\paragraph*{}

This is another principle that has been taken as significant in the search for QG (championed, most notably, within string theory). We claim, however, that mathematical consistency holds a similar status to UV completion. Supposing that the framework of QFT is considered mathematically inconsistent because it yields theories that are not finite ``out of the box'', our arguments largely carry over here, too. If one does not worry about the mathematical (in)consistency of the QFTs of the standard model because of an assumption that these theories are not actually fundamental (i.e., there is physics at higher energy scales that ``explains'' this ``low-energy'' behaviour)\footnote{This relates to a well-known debate between \citet{Wallace2006, Wallace2011} and \citet{Fraser2011}.}, then one should not worry about the mathematical inconsistency of QG, unless one assumes that it is a fundamental or final theory (which, we argue, need not be the case (\S\ref{fund})). That the world be such that the theories we describe it with are mathematically consistent is a bold assumption; the strongest basis for making it is appeal to past success. But, as UV completion is an outdated criterion for selecting physical theories in the age of effective field theory (EFT), so too is mathematical consistency.

\paragraph*{}
Sections \S\ref{princ} and \S\ref{primer} contain the background material: \S\ref{princ} identifies and discusses three main roles that a principle may serve in theory development and evaluation: guiding principle, postulate, and criterion of theory acceptance. We emphasise that these are neither exclusive, nor sharply defined. \S\ref{primer} is a concise introduction to the ideas of renormalizability, EFT, and UV-completion. Here, the message is that there are a number of ways in which UV-completion may obtain---not just through renormalizability.

\paragraph*{}
In \S\ref{why}, we evaluate the general philosophical reasons one might expect that QG be UV-complete. These explain why the principle has traditionally been assumed to serve in the various roles that it has in the search for QG. Broadly, they fall into two categories: Firstly, there is the expectation that QG be the \textit{final unified} theory, or that it at least be a \textit{fundamental} theory. In \S\ref{fund}, we argue that, although UV-completion is necessary for both the fundamentality\footnote{Although, as we mention (``UV silence scenario'', p. \pageref{f}) it is possible that a fundamental theory is not UV-complete.} and finality of a theory, QG is not necessarily fundamental, nor final (i.e. a unified ``theory of everything'' (ToE), compared with a fundamental theory that is not a ToE). We discuss several possible scenarios regarding fundamental physics, and point out that QG should not be assumed in any of them to be UV-complete---\textit{except} if you believe that, fundamentally, the world is best described by QG (understood in the minimal sense, of not being a ToE) plus the standard model of particle physics. While certainly possible, this scenario, however, runs counter to one of the other major principles in QG research: unification. The second general motivation for supposing that QG is UV-complete is associated with the ideas of \textit{spacetime discreteness} and \textit{minimal length}. \S\ref{disc} explores the reasons for thinking that these ideas play a role in QG, and finds that they are not completely well-founded.

\paragraph*{}
Having found no strong general motivations for assuming that QG is UV-complete, we turn, in \S\ref{Approaches} to consider the various approaches to QG, to see how the principle is used in practice. We find, firstly, that a group of broadly related approaches---asymptotic safety, causal dynamical triangulation (CDT), and higher derivative approaches---take UV-completion to be a central principle \textit{without} making a final theory claim. However, we also find that, within some of these approaches, UV-completion does not actually need to play such an essential role as it has been assumed to. Instead, the principle could be weakened (i.e., to \textit{approximate} UV-completion) or removed, and the respective approach would still be viable. On the other hand, to the extent that these approaches do take UV-completion to be a major ambition, we argue that they are not well-motivated in this sense---yet, we cannot rule out the possibility that they turn out to be successful nevertheless. Secondly, we find that within many other approaches, hints at UV-completion are implied (rather than being postulated forthright), and are viewed by their adherents as non-empirical evidence for the correctness of their respective research programs. In these cases, our message is that these approaches are too quick in doing so.

\paragraph*{}
While the preceding sections establish that UV-completion is not well-motivated as a principle of QG, and cannot be used as a criterion of theory selection, the possibility of it serving usefully as a guiding principle is left open. The final section (\S\ref{whyno}), lists several cautions for those intent on using UV-completion to navigate the murky terrain of QG.

\subsection{Principles in theory development and evaluation}\label{princ}
We may roughly distinguish between three different roles of principles in theory development and evaluation: (a) guiding principle; (b) postulate; (c) criterion of theory acceptance. A \textit{guiding principle} is primarily heuristic---it may aid in the discovery of the theory by leading to new insights, but not actually be retained in the resulting theory. On the other hand, a \textit{postulate} is taken to be a pillar of the theory. Finally, for a principle to serve as a \textit{criterion of theory selection} means that a theory should not be accepted if it is somehow incompatible with the principle (unless there is strong evidence in favour of the theory, and/or the principle is shown to be violated under the relevant conditions). The distinction between these different roles is not sharp. Whether a principle serves in role (a) or (b) may only be discernible after the fact---since both begin by being ``postulated'', and then having their consequences in combination with different principles and facts explored, but (b) survives in the theory, while (a) may not (indeed, it may be found to be inadequate or incorrect). An example is Mach's principle in Einstein's search for GR: it was assumed as a postulate of sorts, but then discovered to be too strong. Yet it is still a principle that aided in the theory's discovery. The relationship between (a-b) and (c) is also debatable: For a principle to serve as a criterion of theory selection, it may or may not be also assumed as an ``ingredient'' in the theory. We are sceptical of there being a neat distinction between the ``context of discovery'' and the ``context of justification'' in the case of QG. Principles play a role in driving the search for the theory, the various approaches are built from different (combinations of) principles, and each approach is to be evaluated with respect to principles (as well as empirical evidence). Regardless of how (or where) they feature in this process, it is integral that each of the principles involved is well-motivated \citep{CrowtherRickles}.

\section{Primer on UV-completion, renormalizability, and all that}\label{primer}

Before beginning our appraisal of the role of UV-completion as a principle of QG, we present the (known) ways in which UV-completion may obtain. UV-completion is usually considered in the context of QFT, where it is taken to be synonymous with \textit{renormalizability}. Yet the idea of UV-completion is certainly not restricted to QFT, and there are other ways in which a theory may be UV-complete, apart from being renormalizable. Here, we draw a more general picture of UV-completion. Regarding QFTs, however, we stress (a) the superiority of the fixed-point, non-perturbative criterion of renormalizability over the perturbative one, and that (b) a QFT may be UV-complete without being perturbatively or fixed-point renormalizable---i.e., UV-completion by cutoff, classicalization, or UV/IR-correspondence.

\subsection{Renormalizability and UV-completion}\label{renuv}

In QFT, renormalization is a procedure used to deal with divergences that arise in the theories. These stem from the heart of the formulation of QFT: being a synthesis of quantum mechanics and special relativity. Basically, the combination of Heisenberg's uncertainty principle with the energy-momentum relation, together with the treatment of quantum fields as local (point-based) operators, means that the theory implies the existence of an infinite number of multi-particle interactions.\footnote{This also depends on the type of interaction and the dimension of the momentum space.} A calculation of a particular interaction in perturbative QFT involves a summation over all possible intermediate interactions (perturbative series). These interactions are standardly ordered in terms of loop number. The problem now is that every loop can in principle involve an arbitrarily high exchange of momenta. As the appropriate consideration of a loop term must include integration over all possible momenta, a loop term might thus simply turn out to be divergent — unless further action is taken.\footnote{From the perturbative point of view, the divergences renormalization is concerned with already occur in the single terms in the perturbative series. That the perturbative series as a whole is generically divergent and at most an asymptotic series, is a different, independent issue special to the perturbative approach (cf. \citet{Butterfield}). We have to thank an anonymous referee for pressing us on this distinction.}


\paragraph*{}
The procedure for this (perturbative renormalization) has two steps: First, the infinite integrand in question is split up into an infinite and a finite part (regularization). One way of doing this is by imposing a momentum-/energy-cutoff such that the finite part of the series includes a description of all momentum-energy scales below that of the cutoff, and the infinite part, a description of all momentum-energy scales above it.
The actual physics will in fact always be independent of the means of regularization (cf. \citet{Butterfield}, \citet{Zee}). In the second step, the theory is renormalized by incorporating the infinite part into coupling constants (extra terms that are added expressly for this purpose). This re-parametrisation is complete once the values of the constants have been set using empirical input. Traditionally, if only a finite number of terms have to be introduced to absorb the divergences, the theory is declared \textit{renormalizable}, and it is \textit{prima facie} predictive to arbitrarily high energies.\footnote{There are caveats to the actual UV-completeness, e.g., Landau poles, discussed below.} If an infinite number of terms is required, the theory is non-renormalizable, and---traditionally---non-renormalizable theories were dismissed as \textit{unphysical} on the grounds that an infinite number of experiments would need to be conducted to set their parameters and render them predictive.\footnote{In modern QFT, however, the distinction between renormalizable and non-renormalizable is made at the level of the terms themselves, rather than at the theory-level.} Renormalizable theories were found not only to be predictive, but to be successful, and thus renormalizability became a criterion of theory selection \citep[see, e.g.,][]{Weinberg}.

\paragraph*{} 
A standard, heuristic way of determining whether or not a theory is perturbatively renormalizable is through the \textit{power-counting method}, which considers the momentum scaling behaviour of terms in the perturbative expansion.\footnote{See \citep[cf.][\S 10.1]{PeskinSchroeder} for cases in which the divergence of a term is actually less or more severe than expected from the power-counting criterion.} More precisely, the classification is done by looking at the superficial divergence, $D$, the difference between the power of the momentum in the numerator and the power of momentum in the denominator of a term. For $D>0$, the term is divergent. Significantly, the power-counting criterion states that quantum GR is perturbatively non-renormalizable \citep[see, e.g.,][\S 2.2.2]{KieferBook}.

\paragraph*{}
The development of the effective field theory (EFT) framework and interpretation irrevocably changed this picture. Non-renormalizable theories are considered EFTs, valid far below a certain high-energy scale cutoff, and thus evade the problem of infinitely many loop contributions.\footnote{Note that the higher the loop order, the higher the energy scale; So, if there is an energy cutoff, there is also a maximal loop order contributing.} Non-renormalizable theories can in this sense be thought of as ``effectively perturbatively renormalizable". In other words, they are predictive---and also incredibly successful---so long as we work at relatively low-energy scales. In the context of the standard model, perturbatively non-renormalizable EFTs are typically more useful for making predictions than the underlying (perturbatively renormalizable) theories that they stem from. If we are just interested in the physics at a particular energy scale, then using an EFT is preferable to the full high-energy theory because it captures the relevant physics, using the appropriate degrees of freedom for that particular energy scale \citep{Schwartz,Bain2013}. 

\paragraph*{}
With the advent of EFT came an obscurement in our determination of any theory as potentially fundamental (as defined in \S\ref{fund}). Even if a theory is perturbatively renormalizable, we can never know whether or not there are actually more interactions that become relevant at higher energy scales, and which would then need to be incorporated as additional terms in a new theory. This observation suggests the possibility of a tower of EFTs, each valid at a particular energy scale. Such a tower could either rise to infinitely high-energy scales, or break down at some scale. The latter scenario may occur if QFT itself ceases to be valid at some scale, or it may be that a final, renormalizable high-energy QFT is found---although, we would not be able to tell that this is the case simply from the perspective of that theory. Instead, we would need external evidence for recognising that we have reached the end.

\paragraph*{}
An important concept underlying the framework (and philosophy) of EFT is the \textit{renormalization group} (RG) \textit{flow}. Roughly: because the coupling constants in an EFT at any given scale include the relevant effects of the interactions that occur beyond that scale, the values of these constants will change depending on the scale at which the theory is being used---as more, fewer, or different interactions are taken into account  \citep[see, e.g.,][\S 3.3]{Crowther2016}. The scale dependence of the coupling constants (``running of the couplings") that characterise a particular theory, is described by the \textit{RG equations}. Typically, the running of the couplings is depicted as a flow (i.e., the movement of a point representing a theory) in theory space (the infinite space spanned by all possible couplings). 

\paragraph*{}
It may happen that the RG-flow of a particular theory (in theory-space) hits a \textit{fixed-point} as it flows towards the UV. The couplings in such a theory will thus not change after some high-energy scale. Theories which lie on RG-flow trajectories towards a fixed-point are \textit{non-perturbatively renormalizable} (i.e., they are renormalizable outside of the setting of perturbation theory). A \textit{Gaussian} fixed-point is one where all couplings are zero, and a theory which features one is \textit{asymptotically free}; e.g., quantum chromodynamics (QCD)---the QFT of the strong interaction---runs into a Gaussian fixed-point in the UV. A theory whose RG-flow runs into a \textit{generalised} UV fixed-point---a point where the coupling constants are finite---is called \textit{asymptotically safe}. One of the arguments that QG is UV-complete is the claim that it is asymptotically safe (\S\ref{asy}).

\paragraph*{}
The fixed-point criterion avoids one of the big problems with the perturbative renormalizability criterion, which is its failure to account for \textit{Landau-poles}. These are infinities that can occur in poles in a perturbative theory, even if the theory is supposedly perturbatively renormalizable. A notable example is quantum electrodynamics (QED), which has a Landau-pole at the unimaginably high-energy of $10^{286}$ eV (compare the Planck energy scale of $10^{28}$ eV!). QED is thus incomplete (this worry is often ignored, since researchers typically expect that QED will be incorporated into a unified theory---in fact, the Landau-pole problem may be viewed as for an argument for unification). Theories which are asymptotically safe, such as QCD, however, do not suffer from this issue.

\paragraph*{}
We conclude that while renormalizability does indicate that a theory is UV-complete, it cannot evade the worry caused by the obfuscation of high-energy physics (that is due to the nature of the framework of QFT). It is possible that a renormalizable theory is not actually fundamental---additional evidence is required for the claim that the theory is actually valid at arbitrarily high-energy scales. We now present other forms of UV-completion (which are not free from the epistemic worry, either): their existence demonstrates that renormalizability is not a necessary condition for UV-completion in the context of QFT--albeit that non-renormalizable theories are candidates for UV completion.

\subsection{Other forms of UV-completion}

\subsubsection{UV cutoff}\label{uvc} Another straightforward way of obtaining UV-completion is by imposing a UV-cutoff. However, with a \textit{brute} UV-cutoff comes a direct violation of Lorentz invariance. A brute UV-cutoff amounts to a brutely imposed fundamental length, which imparts a lattice structure upon spacetime; this lattice structure, in turn, introduces a privileged reference frame (the frame in which the lattice is uniform). So, if a UV cutoff is to be adopted, we have to either give up Lorentz invariance (which is problematic since violations of Lorentz invariance at high-energies have been almost ruled out \citep[cf.][]{Liberati}), or smooth forms of a UV-cutoff have to be found. Loop quantum gravity (LQG, \S\ref{LQG}) and causal set theory (\S\ref{CST}) take the latter route.

\subsubsection{UV/IR-correspondence}

UV/IR correspondence denotes a scenario in which high-energy physics (UV physics) looks again like low-energy physics (IR physics). The physicist, Dvali  (and collaborators, see \cite{Dvali}, \cite{Dvali2}) for instance claims that theories of certain kinds are either UV complete by renormalization or by behaving at high energies like classical theories in the ``IR" again — which means that there is an effective high-energy limit to these theories. (From a perturbative point of view, low-energy scattering is classical, as quantum corrections only become relevant towards higher energies.)

\paragraph*{}

In a similar fashion, UV/IR-correspondence \textit{might} also be how string theory could obtain UV-completion: The extended strings, when probed at high energies, correspond to strings at lower energies. This form of UR/IR correspondence cannot be had in a conventional, point-based field theory, as it builds on the duality between strings of length $R$ and strings of length $1/R$ (T-duality)---which requires the basic constituents to have an extension. Yet, because T-duality is not not known for all the objects in the theory, and holds between sectors of the overarching M-theory that are unlikely to sufficiently characterise M-theory, it seems questionable that string theory will turn out to be UV-complete via this UV/IR correspondence.

\section{Why should QG be UV-complete?}\label{why}
There are three general types of considerations that are together responsible for the belief---held by many researchers---that QG is UV-complete.\footnote{For example, there are several different approaches to QG that claim the non-perturbative renormalizability of quantum GR, including asymptotic safety and causal dynamical triangulation. Particular references, as well as more specific motivations for the belief are addressed in the discussions of the individual approaches, \S\ref{Approaches}.} The first of these is the agglomeration of intuitions surrounding the notions of \textit{fundamentality} and a \textit{final theory} (\S\ref{fund}). Others are associated with the ideas of \textit{spacetime discreteness} and \textit{minimal length} (\S\ref{disc}). Related to both of these issues is the \textit{non-renormalizability of gravity} (\S \ref{renuv}; \S\ref{fund}). We find each of these motivations to be problematic. 

\subsection{UV-completion and fundamentality}\label{fund}
The ideas of UV-completion, fundamentality, unification and ``final theory'' are related---exactly how they are connected, however, is typically not fully appreciated. Instead, nebulous (yet well-entrenched) beliefs abound, and these mingle with  retained memories of ``physics past''---remnants from a time when renormalizability was a sure guide to correct physical theories\footnote{Of course, we all know better now, but old principles die hard.}---to form a web of intuitions in which we may have become stuck. Here, we attempt to pull these silver threads apart. 
\paragraph*{}
To begin with, consider what it means for a theory to be \textit{fundamental}: the idea is that it is somehow \textit{basic}. So, a fundamental theory is something like \textit{the most basic description of a given system or phenomenon}. Although not necessary, it is standard to associate this with shorter length scales (higher energy scales), and the idea of describing ``the smallest component parts'' of a system.\footnote{\label{correct}It is the authors' view that speaking of higher-energy theories as ``more fundamental'' is unfortunate in that it connotes that these theories are somehow preferable to their lower-energy counterparts. Instead, we believe that the correct or fundamental description of a system is scale-dependent. Nevertheless, we just work with the standard understanding of fundamentality here.} This owes to the direction of influence described by the RG-flow: the low-energy system depends on the high-energy physics, but not vice-versa. There is also a tendency to think of a fundamental theory as being one that cannot be derived from any other, but what this amounts to is unclear without an account of ``derivation''. Certainly, a fundamental theory cannot be derived from any ``more basic'' theory describing the same system, because none exist (by definition).

\paragraph*{}
Now, consider what a fundamental theory is \textit{not}. It is not necessarily a theory from which all others can be derived (even if you believe there could be such a thing). Neither is it necessarily capable of serving as an ultimate basis for reduction or supervenience. The reason for these negative assertions is that a fundamental theory is not necessarily \textit{lone}. Here, we take as definition that a fundamental theory describes the most basic physics of a particular system or type of interaction (i.e. no ``smaller'' physics beyond)---in this sense, the theory is \textit{final}, but we reserve the designation of ``final theory'' to refer to a \textit{single, unified theory of everything} (ToE) instead. On the other hand, there may be a \textit{set} of fundamental theories, each describing different aspects of the world. This set, taken as a whole, is what could potentially serve as the bedrock (in the absence of a ToE). What is necessarily solo, however, is a ToE. By definition, such a theory is one from which all others could or would (in principle) emerge. A ToE must also be fundamental; and it is typically assumed to describe the ``smallest things'' in the universe.\footnote{Again, this assumption is not necessary.}

\paragraph*{}
It is thus not difficult to see how these concepts have become tangled up with that of UV-completion. A theory that is UV-complete says that the physics it describes is fundamental: it claims that the objects it applies to have no ``smaller component parts'' (note that these claims of fundamentality are made only \textit{by the theory itself}). Yet, a UV-complete theory is not necessarily a ToE. There may be more than one theory that is UV-complete: hence, a UV-complete theory does not necessarily describe \textit{all} of the high-energy physics. Importantly, although it claims to be valid to arbitrarily high-energy scales, a UV-complete theory is not necessarily fundamental---it does not exclude the possibility of there being a theory beyond. However, to discover a theory beyond, we would need to ``step outside'' of the UV-complete theory under consideration. For example, consider QCD, which, as mentioned above (\S\ref{renuv}), is UV-complete because it possesses a Gaussian UV fixed-point. QCD does not describe all high-energy physics (there are more interactions than just quark-gluon ones), but it may or may not be fundamental.\footnote{Also, recall that most of the theories of classical physics are UV-complete, being well-behaved to arbitrary high energies. Yet, these are not considered to be fundamental.} It could be the case that QCD emerges as an approximation to a more basic (yet not necessarily a ToE) theory beyond---perhaps a ToE such as string theory---or it could be the case that QCD, along with the rest of the standard model, plus QG (understood in the narrow sense, as not including condition ToE) together form the most basic description of the world.

\begin{figure}[!ht]\label{Flowchart}
  \centering
  
      \includegraphics[width=1\textwidth]{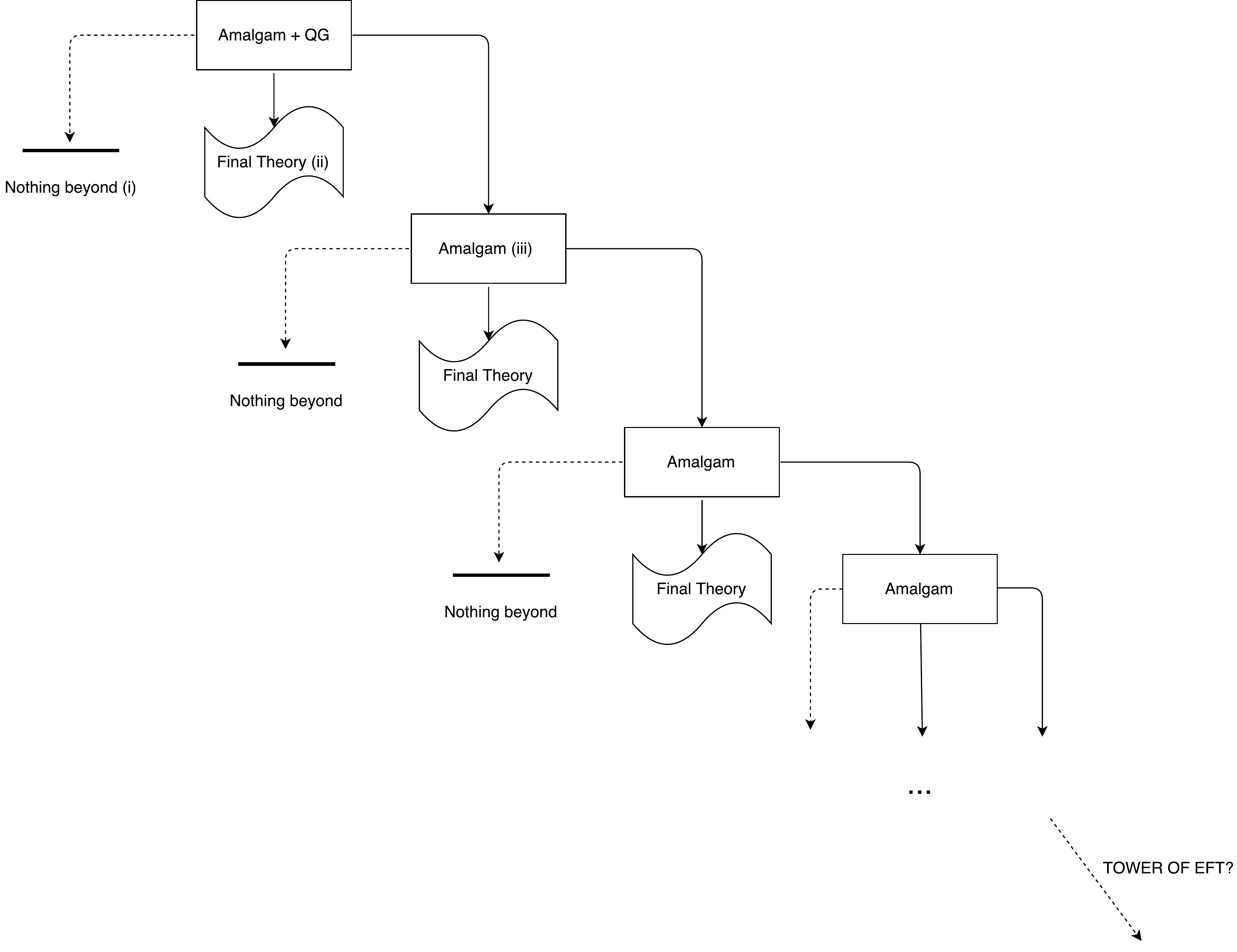}
  \caption{\textit{Options on the road towards higher-energy physics:} QG (as not ToE) is part of the non-unified set of theories ``amalgam'', which is either fundamental (i); emergent from a ToE (ii); or emergent from another amalgam (iii). This is true of any ``level'' described by an amalgam. It is only in option (i) that QG is necessarily -- Discounting the UV-silence scenario -- UV-complete. Solid arrows represent some ``reduces to'' relation (broadly construed; and the reverse direction signals ``emerges from'' or ``arises from'', broadly construed); dashed lines signal the possibility that a given amalgam is the fundamental description.}
\end{figure}

\paragraph*{}
At the first level, as illustrated in Fig. \ref{Flowchart}, we may thus distinguish between two options: either the amalgamation of QG (as not ToE)\footnote{If one includes the criterion that QG be a ToE, then QG is the ToE in (ii), rather than part of the amalgam (i).} plus the standard model (i.e., a non-unified set of theories) are together fundamental (i), or they are not. If they are not, then (some of) these theories are emergent, and they can either emerge from a ToE (ii), or they emerge from another amalgam (iii). And so on. Ultimately, there are three possible scenarios: (1) some non-unified set of theories is fundamental, (2) there is a ToE, or (3) there is no fundamental level, but a never-ending tower of theories.

\paragraph*{}
A few comments are required regarding this first scenario (i), according to which there is no ToE, but rather an amalgamation of several fundamental theories. Many researchers would be loath to accept that this is indeed a possibility---the disunified nature of the amalgam would perhaps dissuade people from believing it fundamental\footnote{Other concerns, such as arbitrariness of the values of the parameters within each theory, the lack of explanatory power, etc. would also contribute to this thought.}, and such researchers would push on in the search for a more unified theory beyond. This would be the case in spite of all the theories in the amalgam being UV-complete, and thus, apparently fundamental.\footnote{One may wish to distinguish between theories that are UV-complete through renormalization, and those that are UV-complete in other ways, and argue that a theory that depends on renormalization cannot be fundamental. One might also wish to argue that no QFT can be fundamental. We discuss these views next.} Since we do not know whether or not there must be a ToE, we seem to be left with two options: If you are OK with an amalgam, then you should search for a UV-complete theory of QG. But, if one does not accept that nature is fundamentally best described by a collection of theories perhaps stitched-up by Dr. Frankenstein, or you are agnostic, then you should not be restricting your search to a UV-complete theory.\footnote{We here in Geneva are proudly celebrating Frankenstein's 200th birthday this year.}

\paragraph*{}
Briefly, then, we should consider whether physics should create such a monster: Can we accept a theory that alludes to physics beyond itself? Actually, there are two questions here; One is whether we can be content with a theory that is not UV-complete---or whether it is our responsibility as scientists to give life to only UV-complete creatures. We maintain that it would be preferable to accept a working, successful non-UV-complete theory \textit{as being the correct theory of QG}, rather than rejecting it in favour of a UV-complete theory that may not be found (of course, you could also accept the non-UV-complete theory as QG, and still continue to search for another theory beyond, if you are so inclined). The other question is whether we can accept a theory if it requires renormalization to be well-defined physically, or whether it is our duty to construct a neat physical theory that fulfills its intended purpose without requiring additional mathematical surgery. Again, if there is a choice between a successful renormalizable theory (UV-completion by renormalizability) in hand, and a hypothetical directly UV-complete (no need of renormalization) one in the (not very promising looking) bush, we think the choice is obvious enough. This is the case even if we are considering fundamental theories.

\paragraph*{}
Also---importantly---for both these questions there is a stronger philosophical motivation for not taking pains in looking for a UV-complete theory of QG. This is the incomprehensible nature of what it is that you are claiming to seek: A theory valid to arbitrarily high-energy scales. It is hubris that compels one to fly so high---albeit, this hubris comes from the incredible success of physics so far. Nevertheless, no matter how successful our current theories, we have no guarantee that they will yield a correct description of the extreme high-energy---especially given (or in spite of) the variety of ways we can stretch and manipulate these theories to fit these regimes (\S\ref{Approaches}).

\paragraph*{}
Still, owing to some issues with QFT (including some related to the necessity of renormalization), there is a sentiment that the framework is not fundamental. If this is so, then the first case described above (the amalgam of QG and the standard model as fundamental) is not possible, and, although QG may be fundamental (if it is UV-complete and not a QFT), the standard model must be emergent. But this does not mean that there is a ToE beyond---rather, it could be that we have an amalgam of several \textit{other} fundamental theories (which may or may not include QG), and it is from these that the standard model (and QG, if it is not among the fundamental theories), may emerge. Or, it could be that there is no fundamental description, but rather a never-ending \textit{tower of theories}.\footnote{Cf. \citet{CaoSchweber} and \citet[][\S 3.5--3.8]{Crowther2016}.} Again, in any of these scenarios except for (i), we should not assume that QG must be UV-complete.\footnote{Here---to emphasise---we are remaining open as to whether or not QG is a ToE.} But, as well as being unpalatable for the reasons given above, this scenario stands in conflict with one of the strongest motivations for QG: unification. Thus, it seems, seeking a UV-complete theory of QG---\textit{qua} QG---means favouring the principle of UV-completion at the expense of the principle of unification. As we discuss further in \S\ref{whyno}, UV-completion may come into conflict with other desirable, and (presumably) viable principles of QG as well: We have already mentioned the case of Lorentz invariance (\S\ref{uvc}, also \S\ref{CDT}), but another example is unitarity (\S\ref{hda}). 

\paragraph*{}
In general: being UV-complete is considered to be necessary, but not sufficient, for a theory to be fundamental; and UV-completion is necessary, but not sufficient, in order for a theory to be ``final'' (i.e., a ToE). Yet, even this might be questioned: if UV-completion was not necessary for fundamentality, it would mean that the world is just not amenable to scientific description at extremely high-energy scales.\label{f} This ``UV silence scenario'' is a possibility worth mentioning, although we do not take it seriously here.

\paragraph*{}
Regarding QG: the theory is not necessarily UV-complete,\footnote{As we shall see when discussing the particular approaches (\S\ref{Approaches}), however, some of these are necessarily UV-complete.} nor must it be fundamental, or final (ToE). Yet, the desire for these things has historically featured as part of the motivation for QG. We have shown that these ideas are distinct, and should be separated from QG. One might wonder if, nevertheless, these objectives might usefully serve \textit{heuristically} as guiding principles in the search for a theory---and, because UV-completion is necessary both for fundamentality and finality (i.e., ToE), this would lead you to begin by striving for a theory that is UV-complete. Certainly, this is true---but it is also bad practice to embark with a particular principle in mind without reasonable motivation for it. In the past, when renormalizability was used as a guide to physical theories, there was reasonable justification for assuming UV-completion, since renormalizable theories were found to be successful; but this time has passed, and non-renormalizable theories are now recognised as successful themselves. We are not claiming that UV-completion must be forsaken; instead, we maintain that the search for QG should be broadened to seriously include the possibility that QG is not UV-complete.

\paragraph*{}
The outmoded attitude towards non-renormalizable theories lies behind one of the traditional objections to quantum GR: the theory's non-renormalizability. However, as mentioned above, it may be that the apparent non-renormalizability is just due to the inapplicability of perturbation theory, rather than any actual problems with the theory at high energies. It is possible that the theory---which is typically regarded as an approximation to QG---is UV-complete (perhaps possessing a UV fixed-point, as in the asymptotic safety scenario). In such a case it is even possible (however unlikely) that this theory turns out to be QG---if it fulfills the set of criteria that are taken to define QG (such as describing the physics where GR and quantum theory intersect, including at the Planck scale).\footnote{Although we may worry that perhaps a UV-complete theory is necessary to probe singularities.} But even if the theory is actually non-renormalizable, it is still useful and predictive: reproducing the results of QG at accessible energy scales.

\paragraph*{}

\citet{Hossenfelder} reviews the history and basis for the idea of a minimal length scale in QG; including Werner Heisenberg's belief (in the 1930s and 40s) in a minimal length, based on problems he saw with the (then incomplete) theory of the fundamental interactions, particularly 4-Fermi theory. Yet, the minimal length that Heisenberg envisaged (of the order $100$ fm) was not discovered; instead QED was developed, the strong and electroweak interactions were found, and 4-Fermi theory of  $\beta$-decay was determined to be non-renormalizable, and not fundamental. Hossenfelder (2013, p. 10) thus poses the question, 
\begin{quote}
	``So we have to ask then if we might be making the same mistake as Heisenberg, in that we falsely interpret the failure of general relativity to extend beyond the Planck scale as the occurrence of a fundamentally finite resolution of structures, rather than just the limit beyond which we have to look for a new theory that will allow us to resolve smaller distances still?''
\end{quote}
\citet{Hossenfelder} argues that it is not a mistake to believe in the fundamentality of the Planck scale---citing various arguments for the existence of a minimal length (which we consider in \S\ref{disc}) that, she says, feed ``the hopes that we are working on unveiling the last and final Russian doll'' (p. 10). We argue here that researchers have been too quick to infer that QG is the theory that will realise their hopes of finality (or even fundamentality). Instead, the prudent---and historically justified---second possibility (of just requiring a new theory) is more reasonable. Yet, we also wish to point out that neither option may be correct: Instead, it is possible that the theory remains useful beyond the Planck scale  \citep[as in, e.g.,][]{Percacci2010}. 

\subsection{UV-completion and minimal length}\label{disc}
As we see when discussing the different approaches to QG, some of them get a UV-complete theory by featuring a \textit{minimal length}. This is understood most generally as an operational limit, meaning nothing smaller can be probed. It may obtain from an extended fundamental probe (e.g. string theory, \S\ref{string}), from spacetime discreteness (e.g., causal set theory, \S\ref{CST}, LQG \S\ref{LQG}), or in some other way (e.g., ill-defined geometry, or an uncertainty principle). Outside of the context of particular approaches, the general heuristic arguments for minimal length involve aspects of both GR and quantum theory. 
The determination of the Planck length (time, and mass) using dimensional analysis, as well as thought experiments involving ``Heisenberg's microscope'' in the presence of gravity.\footnote{Cf. \citet[][\S5.5-5.6]{Hagar}, \citet[][\S3.1]{Hossenfelder}.} Such considerations have sometimes been taken to suggest that geometry is ill-defined, or ``fuzzy'', at the Planck scale---perhaps a ``quantum foam'' of microscopic, rapidly-evaporating black holes. More generally, many researchers believe that these arguments establish the existence of a fundamental operational limit on how far we can probe physics at small distances. However, this places too much weight on the extrapolation of GR and quantum theory into distant realms. Semiclassical theories are certainly useful for exploring QG phenomenology, and play an invaluable role in the development of QG. But, we stress the need for caution in trusting any claims regarding the ``end of physics'' that come about from manipulating and stretching GR and QFT into inaccessible energy scales, particularly the Planck scale.\footnote{\citet{Wuthrich} also expresses this opinion.}  

\paragraph*{}
A minimal length also comes about from various attempts to quantize the gravitational field. An example is LQG (\S\ref{LQG}), where spacetime discreteness arises in the covariant approach to quantizing geometry (where the quanta are ``fuzzy tetrahedra'', due to their being characterised by non-commuting operators) as well as in the canonical approach to LQG (which describes discrete spectra for area and volume operators) \citep{Rovellicovariant, RovelliCanonical}. In the second case, as has been well-publicised, the discreteness is a prediction of the theory. This has been championed as an achievement of the theory---demonstrating consistency with the suggestions from heuristics, and supposedly attaining a sort of mutual justification. This is perhaps not surprising, since these approaches were derived from the same source material. Their support of one another is not evidence of the theory's correctness. And neither is the theory's UV-completeness---whether assumed upfront or derived as a prediction. There is no requirement that QG be UV-complete. Also, we must stress that neither is it necessary that QG be a theory of quantized geometry, as \citet{Wuthrich2005} demonstrates; so, even if it is true that a theory of quantized spacetime is UV-complete, this does not establish that QG is UV-complete. 

\paragraph*{}
Finally, we mention an argument for a minimal length scale in the context of asymptotic safety (\S\ref{asy}). It is particularly interesting because asymptotic safety claims that quantum GR is non-perturbatively renormalizable (possessing a UV fixed-point), and thus eliminates one of the most-cited pieces of evidence for a minimal length. The asymptotic safety scenario claims that quantum GR fundamental, and \citet{Percacci2010} argues that in a fundamental theory, a physical unit of measurement should be defined within the theory, and so will be done using its dimensionful coupling constants. Thus, the unit will be affected by the running of the couplings under RG-flow. Due to the presence of a UV fixed-point, the physical energy---i.e., energy \textit{measured in an appropriate energy unit}---might go to a finite (maximal) value as the bare energy parameter (denoting the theory in theory space) goes to infinity. This finite energy value would represent a fundamental length scale. This holds at least in the special case where only the running of the standard couplings of GR (i.e., the gravitational and cosmological constants) are considered.

\paragraph*{}
 
However, it is possible that GR is not actually asymptotically safe, in spite of the evidence for this scenario. Instead, as \citet{Percacci2010} demonstrate, the theory may only be \textit{approximately} asymptotically safe: it appears to us, low energy observers far from the Planck scale, as if the theory's couplings flow to a fixed-point, but really their trajectory is just ``nearly'' asymptotically safe (the theory may actually run away from the fixed-point at higher energies, rather than into it). In such a scenario, there may be no operational limit to us probing energies higher than the Planck mass--except that this is the scale at which new physics is expected to appear.

\section{UV-completion in different approaches to QG}\label{Approaches}

Pick up any textbook or review article on (any particular approach to) QG, and you are likely to find that somewhere in the beginning, it raises the non-renormalizability of perturbative quantum GR as a problem to be overcome by QG. The article will then proceed to demonstrate how a UV-complete theory may be obtained---and the offending perturbative non-renormalizability thus rendered impotent---in the particular approach being promoted (e.g. \cite{Ambjorn} in the case of CDT, \cite{NiedermaierReuter} regarding asymptotic safety, any textbook on string theory). Thus, one could very easily get the impression that this represents \textit{the} problem of QG. The different approaches to QG can even be categorised based on how they attempt to establish UV-completion:

\begin{itemize}
\item
At the perturbative level, through modifications of GR (e.g. The ``higher derivative'' approaches, which employ additional propagators; supergravity, which uses supersymmetry);
\item
At the non-perturbative level (e.g. Asymptotic safety, which features a UV fixed-point; CDT, which features a continuous limit of path integral over regularized geometries); or
\item
By moving to a higher dimensional (generalised) QFT (e.g., String theory\footnote{Perturbative string theory is arguably a higher-dimensional generalisation of perturbative QFT when seen in the world-line picture. (Feynman's world-line picture is simply generalised to a world-sheet picture).}, which features extended basic objects).
\end{itemize}

Yet---and we must emphasise this (cf. \S\ref{why})---the perturbative non-renormalizability of the theory is not even necessarily a problem, so long as the theory covers the required regimes (including Planck scale physics).\footnote{In the standard effective field theory view on QFT, higher energy modes---suppressed in the effective field theory view---are expected to become relevant already below the cutoff. As the cutoff energy in perturbative quantum GR } \textit{There is no requirement that QG be UV-complete}.

\paragraph*{}
Briefly, we also wish to comment on the status of perturbative versus non-perturbative approaches. Canonical approaches to QG are typically seen as attempting to establish UV-completion through their employment of non-perturbative methods (i.e. one might think to list them alongside asymptotic safety and CDT in the categorisation above). But canonical approaches do not necessarily seek UV-completion through their employment of non-perturbative methods---the use of these methods may already offer advantages over perturbative ones, without requiring UV-completion. A (heuristic) argument that a non-perturbative approach to QG provides additional insights over the standard perturbative approach is given in \citet{Ashtekar}. It shows that non-perturbative approaches can evade \textit{certain} divergences that perturbative ones cannot. It does not, however, establish that a non-perturbative account of QG should be UV-complete.

\paragraph*{}
We now turn to explore the role of UV-completion in different approaches to QG. Here, we ask whether UV-completion is used as a guiding principle or as a criterion of theory choice within each specific approach \textit{in practice}. In cases where it is relevant, we also consider whether UV-completion \textit{should} be used as a guiding principle or criterion: i.e., we attempt to discover how integral UV-completion actually is to each approach (given the particular aims and methods of the respective approach).
\subsection{String theory}\label{string} 

String theory originally arose from an unsuccessful theory of hadrons, but evolved into a candidate ToE. Two of the major discoveries that contributed to this progress were: that (1) UV-completion could (most likely) be obtained through the postulation of extended objects as the basic entities (e.g., strings, and higher dimensional objects called branes), and that (2) the graviton was among the excitations in the closed string's spectrum. Thus, string theory has been promoted as a candidate for QG, as well as a candidate ToE (since the other particles of the standard model also appear as excitations of the string).

\paragraph*{}
As a candidate ToE, string theory must be UV-complete,\footnote{Discounting the ``UV silence scenario'' described on p. \pageref{f}.} and the extendedness of the theory's basic entities is very likely to ensure that it is; although no proof has yet been found that string theory is UV-complete (see \citet[][\S 7.2]{Hagar}, \citet[][\S 1]{Dawid}). Often, textbooks motivate string theory by drawing an analogy between the perturbative non-renormalizability of GR and that of 4-Fermi theory (which was revealed to be the effective limit of the renormalizable electroweak theory): The non-renormalizability of GR indicates that one \textit{must} look for a renormalizable theory that has quantum GR as an EFT, and string theory is exactly this kind of theory. Thus, the alleged UV-completeness of string theory is presented as one of its selling points (i.e., as a criterion of justification).

\paragraph*{}
String theory is remarkable for many reasons, not least because it gives rise (among other things) to the graviton, and contains rich, promising structures. If---contrary to common expectations and reassurances from the string community (see \citet{Dawid})---string theory turned out to be UV incomplete, it is not clear what problems this would cause, \textit{apart} from its having to renounce its claim of being the ToE.\footnote{If we were to accept the UV silence scenario, however, then string theory could still be the ToE, even without being UV-complete.} The theory has enough interesting features to ensure that it would not be discarded. It would still be useful, despite not making predictions at arbitrarily high-energy scales; and it could still be the correct theory of QG.

\paragraph*{}
The fixation with string theory's UV-completeness may be dulled, too, by recalling the scepticism issue inherent to modern high-energy physics (\S\ref{renuv}): Even if a theory is renormalizable (or otherwise UV-complete), this is no guarantee that it is the correct theory at all high-energy scales (new physics might still exist at some scale). UV-completion, and renormalizability are formal properties of theories.

\subsection{Asymptotic safety}\label{asy} Inspired by the asymptotic freedom of QCD, the asymptotic safety scenario builds on the assumption that GR runs into a UV fixed-point such that the couplings remain finite. (Unlike in the asymptotic freedom scenario in QCD, the fixed point is conjectured to be non-Gaussian.)\footnote{More precisely, \citet{NiedermaierReuter} distinguish between \textit{weak renormalizability} in the UV, i.e., (1) the number of irrelevant couplings (couplings which decrease towards the fixed-point) is finite and (2) the number of relevant couplings (couplings which increase towards the fixed-point) may be infinite but their values remain finite, and \textit{quasi-renormalizability}, i.e., weak-renormalizability with the additional requirement that the number of relevant couplings decreases to a finite amount in the UV (due, e.g., to symmetries). It is the latter sense of renormalizability we have in mind here since only it allows for predictability towards the UV.} In their review, \citet{NiedermaierReuter} stress the fundamentality of the gravitational degrees of freedom but also the dependence of the action on the energy scale, since the RG-flow might add new terms to the Einstein-Hilbert action while changing the gravitational and the cosmological constants.\footnote{This may pose the danger of violating unitarity, similiar to the case regarding the higher derivative approaches (\S\ref{hda}).}

\paragraph*{}
The idea of UV-completion (by a UV fixed-point) is the central guiding principle of asymptotic safety. Meanwhile, evidence for the existence of a fixed-point has been found under the truncation of theory space and the restriction of theory space through the assumption of certain symmetries \citep{NiedermaierReuter, Bain}. Yet, asymptotic safety makes some strange predictions, such as the reduction of spacetime dimensions from four in the IR to two in the UV.\footnote{Interestingly, CDT (\S\ref{CDT}) shares some of these predictions, including dimensional reduction and the existence of a UV fixed-point.}  As asymptotic safety only works with UV-completion in the form of a fixed-point criterion, the possibility of \textit{hidden} couplings that become relevant at higher energies (\S\ref{renuv}) still exists. Thereby, the UV-completion, although it might serve as a guiding principle, could ultimately turn out to only hold approximately (if at all). The scenario of \citet{Percacci2010} (mentioned above, \S\ref{disc}), is one possibility. Thereby, asymptotic safety is potentially compatible with the idea of \textit{emergent gravity} (\S\ref{emerg}). 

\subsection{Causal dynamical triangulation}\label{CDT} 

CDT applies the path integral quantization over spacetime geometries. To evade issues of divergences possibly rooted in the perturbative treatment of the path integral, the approach does without expanding the path integral in perturbative terms and builds instead on the idea of first taking discrete geometries as the set of geometries to sum over (as a form of regularization) to apply a limit procedure on them afterwards. As it is committed to a well-defined continuous limit over these geometries\footnote{Note that this is not the same as the semi-classical or classical limit, by which theories of QG attempt to recover GR.} \citep{Ambjorn}, it presupposes UV-completion. The idea that quantum GR will be UV-complete thus seems to be an underlying guiding principle. Yet, it is conceivable that CDT forgo the limiting procedure by which the dependence on regularization is usually removed. This could either be interpreted as imposing a brute UV cutoff for CDT, or taking CDT to be only valid as an EFT.

\subsection{Higher derivative approaches} \label{hda} 
Higher derivative approaches are so-called because they include higher derivative terms of the metric in the Einstein-Hilbert action, in an attempt to \textit{regain} perturbative renormalizability. Thus, UV-completion is taken as a central guiding principle. Yet, this may lead to a conflict with another significant principle in physics: Unitarity. Consider, for instance, the case of adding to the Einstein-Hilbert Lagrangian a correction term proportional to $R^2$: The inclusion of this higher derivative term leads to additional graviton propagators with a momentum dependence of $p^{-4}$ rather than $p^2$. Adding these higher momentum propagators helps in achieving perturbative renormalizability, but one of them allows for tachyons (particles with negative rest mass which can move backwards in time) and thus the violation of unitarity (see \citet{KieferBook}; \citet{Stelle}). Consequently, support for these approaches amounts to favouring the principle of UV-completion over that of unitarity.

\subsection{Supergravity}

Supergravity combines conventional GR with supersymmetric fermionic and bosonic fields by straightforwardly extending the Einstein-Hilbert action through the addition of terms corresponding to these fields. While also being a field of study in its own right, supergravity is notable for being one of the weak-energy limits of string theory (and famously starring in the AdS/CFT correspondence). The introduction of additional symmetries in the action may be motivated as a means of improving the renormalizability of standard perturbative QG---and thus UV-completion could be seen as one of its guiding principles. Divergences are however still expected even for $\mathcal{N}=8$-supergravity---the most symmetrical supergravity in 4-dimensions, and thus presumably the most renormalizable one (cf. \citet{Deser1977}, \citet{Bern}).
Nevertheless, supergravity is worth exploring whether it is UV-complete or not (after all, matter might simply be supersymmetric), and thus UV-completion need not be viewed as its central motivation.

\subsection{Causal set theory}\label{CST} 
Causal set theory is based on the result of Malament's work that the spacetime structure in GR can be split up into (continuous) causal structure and local volume information \citep{Malament1977}. Starting with a \textit{discrete causal set structure}, it is possible to reobtain the local volume information (of a spacetime) by means of a simple counting procedure. Thus, when seen at larger length scales, a discrete causal structure can in principle be used to model a continuous spacetime. The main motivation for choosing a \textit{discrete} causal structure in causal set theory is apparently to facilitate the attainment of local volume information from the causal structure.

\paragraph*{}
Causal set theory is neither a quantum theory nor an equivalent formulation of GR (there is no equivalent to the field equations, for instance). It is a reformulation of a GR-like spacetime structure in terms of a discrete structure, such that at large length scales the ordinary spacetime structure shows up again. What is the motivation for choosing causal set theory over GR? One response is that the sheer possibility of reproducing the spatiotemporal relations of GR \textit{purely} by means of discrete causal structure is worth studying on its own right. We cannot disagree with that. Another answer is that we need to have discreteness as the basis of our theory of spacetime anyway.\footnote{\citet{Henson} and \citet{Reid} go this way. See \citet{Wuthrich} for an appraisal.} But, it is not clear to us why discreteness should already be forced onto our spacetime theory at the \textit{classical} level, and especially in the peculiar way that causal set theory does it.

\paragraph*{}
Still, a stronger answer to this question would be that causal set theory represents an improvement over GR because its fundamentally discrete structure makes it more likely that the corresponding quantum theory will be UV-complete. A discrete structure allows for a UV cutoff, and causal set theory even offers a means of implementing a UV cutoff compatible with Lorentz invariance.\footnote{This is the random ``sprinkling'' procedure, which ensures that no observer is favoured by the introduction of the cutoff structure \citep[cf.][]{Dowker2005}.} If this is the answer given by a particular researcher, it signifies their adoption of UV-completion as a major guiding principle.

\subsection{Canonical QG}\label{CanQG} Apart from the path integral method, canonical quantization is the standard prescription for turning a classical theory into a quantum theory. Nothing in the application of the canonical prescription seems to rest on the idea of UV-completion (although there might be a hope that the non-perturbative quantum GR thus obtained is UV-complete).

\subsection{Loop quantum gravity} \label{LQG} LQG (both in its canonical~\citep{RovelliCanonical} and covariant formulations~\citep{Rovellicovariant}) leads to discrete chunks of spacetime (at least from an operational point of view), which---due to the peculiar nature of these chunks---gives rise to a \textit{Lorentz-invariant} cutoff (UV-completion by cutoff). This is taken by proponents of LQG as strong justification for the approach. There seems no need for UV-completion to serve as a guiding principle for a quantization approach (since it is obtained automatically).

\subsection{Approaches based on alternative gravitational theories}

We cannot go through all alternative theories of gravity. Rather, we limit our considerations to: Brans-Dicke, Horava-Lifshitz and shape dynamics (SD). Within this selection, it turns out that in each of these approaches, UV-completion is noted in some sense and typically presented as strong evidence in favour of the respective approach.\footnote{For Brans-Dicke gravity, see, e.g., \citet{BransDi2} \textit{contra} \citet{BransCite1} \textit{and} \citet{HooftVeltmann}. For Horava-Lifshitz gravity, renormalizability is shown in the power-counting sense, but it is otherwise subject to discussion~cf. \citet{Horava}. As SD shares the same local symmetries in the UV as Horava-Lifshitz gravity, and these symmetries allow for power-counting arguments in favour of the renormalizability of gravity, SD seems to be power-counting renormalizable if Horava-Lifshitz gravity is power-counting renormalizable (cf. \citet{Gryb})}

\paragraph*{}
Yet, UV-completion does not feature as a guiding principle in the construction of alternative theories of gravity. The motivations for Brans-Dicke and SD are classical considerations. For instance, they both take inspiration from (different readings of) Mach's principle, which is, loosely speaking, the idea that spacetime structure should be sufficiently determined by its matter content. More precisely, SD is guided by the idea that a theory of spacetime should do without absolute notions of space and scales. In the case of Horava-Lifshitz gravity, the central motivation comes from the challenge to unify the different concepts of time in QFT and GR. Its main aim is to replace the notion of external time in quantum theory with a conception of time that accords with the lessons of GR.

\subsection{Emergent gravity approaches}\label{emerg}

The most eclectic class of approaches to QG is the one denoted by the title of ``emergent gravity''. The approaches within this category treat GR as an effective theory---meaning that the gravitational interaction is not fundamental, but only describes superficial degrees of freedom.  Some of the approaches stress the thermodynamic-hydrodynamic nature of GR; others try to render GR as a bulk effect from a more basic theory on a holographic surface.\footnote{Hydrodynamic, see, e.g., \citet{Jacobson, Padmanabhan2}; holographic, see, e.g., \citet{Jacobson2015, Verlinde}.} The condensed matter approaches to QG build upon a strong analogy between GR and the low-energy theories of quantum fluids (condensates).\footnote{See, e.g., \citet{Visser, Bain} and \citet{Barcelo}.} From the perspective of all of the emergent gravity approaches, the quantization of gravity is a red herring in regards to the true theory ``beyond'' GR: although quantization will give you a theory of gravitons, or ``quanta of spacetime'', these are (according to emergent gravity) just quasi-particles (i.e. low-energy phenomena produced by collective degrees of freedom), and \textit{not} the fundamental degrees of freedom of the underlying theory. The more important point for us here, though, is that these approaches renounce UV-completion as a guiding principle: gravity is not a fundamental interaction, and so its theory need not (and, in some approaches \textit{should not}) be predictive to arbitrarily high-energy scales.

\paragraph*{}
Interestingly, asymptotic safety might still be used as a guiding principle in these approaches. \textit{Prima facie}, this sounds like a contradiction: asymptotic safety is based on the idea of UV-completion, while the emergent view of gravity does not expect UV-completion. But, as mentioned above (\S\ref{disc},\ref{asy}), \citet{Percacci2010} demonstrates that asymptotic safety---along with the apparent UV-completion that it bestows---may only hold to a first approximation. Hence, we could explain the apparent evidence for a UV fixed-point for gravity, and avail ourselves of any theoretical conveniences that the assumption may provide---and yet still give full credence to the possibility of emergent gravity. \cite{Bain} also argues that emergent gravity (the condensed matter approaches) may leave room for the principle of asymptotic safety (but does so from another perspective).

\section{UV-completion as a guiding principle in QG}\label{whyno}

So far, we have argued that UV-completion is neither necessary, nor well-motivated as a guiding principle of QG; that it should not be adopted as a postulate; and that it cannot be used as a criterion of theory selection. But, some readers may still (as in \S\ref{disc}) object to it being discarded as a guiding principle. Again, the suggestion is that---given that UV-completion has traditionally (and often successfully) featured as a motivation in theory development---it may continue to serve usefully as a heuristic guiding principle, even if we recognise that the theory we are searching for is not necessarily UV-complete, and that UV-completeness is no guarantee of the theory's fundamentality. Also, one might think that even if the principle is not well-motivated, it is still a possible constraint that may help steer our course through a landscape with precious few bounds.

\paragraph*{}
In response, we again note that our arguments in this paper do not mean that UV-completion must be rejected as a guiding principle of QG. Instead, we propose that it may be useful to relax or drop the principle in this particular role. If one nevertheless feels compelled to retain it as an inessential heuristic aspiration, we suggest you do so with caution, taking into account the following lessons from the above investigation:

\begin{itemize}
    \item Motivations for UV-completion are outdated, or based on incorrect beliefs about what QG must be like;
    \item UV-completion illegitimately rules out large class of theories;
    \item UV-completion is an unnaturally restrictive criterion that may result in contrived theories; It may impose features that are otherwise unmotivated, such as dimensional reduction; and,
    \item UV-completion may conflict with other principles of QG that are otherwise (apparently) viable, and in some cases better motivated, e.g., Lorentz invariance, unitarity, unification.
\end{itemize}

\section{Conclusion}
In the search for QG, UV-completion has been taken as an important, yet unevaluated, principle---serving not only as motivational guidance, but as a postulate and as a criterion of theory justification. Physicists working on QG typically view themselves as seeking a fundamental theory rather than an effective one (i.e. a theory whose validity is restricted to certain scales), and thus hold UV-completion as an essential constraint. We have argued, however, that this is misguided. There is no requirement that QG be valid to arbitrarily high-energy scales (or to the shortest length scales), and thus, UV-completion cannot be taken as a criterion of theory acceptance. Instead, the necessary requirement is more modest: that the theory be \textit{UV-better} (than what we have now)---i.e., that it be valid at the Planck scale.\footnote{We thank Jeremy Butterfield for suggesting this slogan.} UV-completion only makes sense as criterion within approaches whose goal is a ToE---yet, most approaches to QG do not have this aim.
\paragraph*{}
In many approaches where UV-completion is obtained consequentially, rather than being postulated up front, it is hailed as non-empirical support for the theory. Yet, the attainment of UV-completion is not evidence for the correctness of the approach, and \textit{especially} not for those approaches that do not aim at a unified theory of everything. This is true even when UV-completion comes through the approach's prediction of a minimal length. One of the problems here is that heuristic arguments for the existence of a minimal length are, in general, treated too strongly---and a strange impasse has been created through the belief that the heuristic arguments for a minimal length, and the particular approaches to QG that predict a minimal length, mutually support each other. 

\paragraph*{}
As a guiding principle, the idea of UV completion is a hangover from the heyday of the discovery of the success of renormalizable QFTs. In our investigation of the particular approaches, we found that UV completion is not actually necessary within several approaches that treat it as a central postulate or guiding principle. Instead, in these approaches, it could be weakened or dispensed with in favour of other principles. 

\paragraph*{}
However, none of this rules out the possibility that QG is UV-complete, nor does it establish that the adoption of UV completion as a guiding principle would necessarily prevent any particular approach from discovering the correct theory. Our main arguments have established, simply, that QG is not necessarily UV-complete, and that UV completion is both poorly-motivated and potentially undesirable as a guiding principle. Instead of imposing upon the inaccessible, it is better to exercise high-energy humility.

\section*{Funding}

Funding for this research was provided by the Swiss National Science Foundation (105212\_165702).

\section*{Acknowledgements}

We first and foremost express our gratitude towards Juliusz Doboszewski for insightful conversations on the topic. Furthermore, we thank Christian Wüthrich, Rasmus Jaksland, Claus Beisbart, Vincent Lam, Marko Vojinovic, Tushar Menon, Augustin Baas, Baptiste Le Bihan, James Read, and audiences in Lausanne, Munich and Oxford for valuable feedback.
We are also grateful for the helpful comments of three anonymous referees.

\bibliography{references.bib}

\begin{thebibliography}{}

\bibitem[Ambjørn et~al., 2014]{Ambjorn}
Ambjørn, J., Görlich, A., Jurkiewicz, J., and Loll, R. (2014).
\newblock {Quantum Gravity via Causal Dynamical Triangulations}.
\newblock In Ashtekar, A. and Petkov, V., editors, {\em Springer Handbook of
  Spacetime}, pages 723--741.

\bibitem[Ashtekar, 1991]{Ashtekar}
Ashtekar, A. (1991).
\newblock {\em Lectures on non-perturbative canonical gravity}, volume~4.
\newblock World Scientific.

\bibitem[Bain, 2013]{Bain2013}
Bain, J. (2013).
\newblock Effective field theories.
\newblock In Batterman, B., editor, {\em The Oxford Handbook of Philosophy of
  Physics}, pages 224--254. Oxford University Press, New York.

\bibitem[Bain, 2014]{Bain}
Bain, J. (2014).
\newblock Three principles of quantum gravity in the condensed matter approach.
\newblock {\em Studies in History and Philosophy of Science Part B: Studies in
  History and Philosophy of Modern Physics}, 46:154--163.

\bibitem[Barcel{\'o} et~al., 2005]{Barcelo}
Barcel{\'o}, C., Liberati, S., Visser, M., et~al. (2005).
\newblock Analogue gravity.
\newblock {\em Living Rev. Rel}, 8(12):214.

\bibitem[Bern, 2002]{Bern}
Bern, Z. (2002).
\newblock Perturbative quantum gravity and its relation to gauge theory.
\newblock {\em Living Rev. Rel}, 5(5).

\bibitem[Butterfield and Bouatta, 2015]{Butterfield}
Butterfield, J. and Bouatta, N. (2015).
\newblock Renormalization for philosophers.
\newblock In {\em Metaphysics in Contemporary Physics}, pages 437--485. Brill.

\bibitem[Cao and Schweber, 1993]{CaoSchweber}
Cao, T.~Y. and Schweber, S.~S. (1993).
\newblock The conceptual foundations and the philosophical aspects of
  renormalization theory.
\newblock {\em Synthese}, 97(1):33--108.

\bibitem[Crowther, 2016]{Crowther2016}
Crowther, K. (2016).
\newblock {\em Effective Spacetime: Understanding Emergence in Effective Field
  Theory and Quantum Gravity}.
\newblock Springer, Heidelberg.

\bibitem[Crowther and Rickles, 2014]{CrowtherRickles}
Crowther, K. and Rickles, D. (2014).
\newblock Introduction: Principles of quantum gravity.
\newblock {\em Studies in History and Philosophy of Science Part B: Studies in
  History and Philosophy of Modern Physics}, 46:135--141.

\bibitem[Dawid, 2013]{Dawid}
Dawid, R. (2013).
\newblock {\em String theory and the scientific method}.
\newblock Cambridge University Press.

\bibitem[Deser et~al., 1977]{Deser1977}
Deser, S., Kay, J., and Stelle, K. (1977).
\newblock Renormalizability properties of supergravity.
\newblock {\em Physical Review Letters}, 38(10):527.

\bibitem[Deser and van Nieuwenhuizen, 1974]{BransCite1}
Deser, S. and van Nieuwenhuizen, P. (1974).
\newblock One-loop divergences of quantized einstein-maxwell fields.
\newblock {\em Physical Review D}, 10(2):401.

\bibitem[Dowker, 2005]{Dowker2005}
Dowker, F. (2005).
\newblock Causal sets and the deep structure of spacetime.
\newblock In Ashtekar, A., editor, {\em 100 Years of Relativity: Space-time
  Structure}, pages 445--467. World Scientific, Singapore.

\bibitem[Dvali, 2011]{Dvali}
Dvali, G. (2011).
\newblock Classicalize or not to classicalize?
\newblock {\em arXiv preprint arXiv:1101.2661}.

\bibitem[Dvali et~al., 2011]{Dvali2}
Dvali, G., Giudice, G.~F., Gomez, C., and Kehagias, A. (2011).
\newblock {UV-completion by classicalization}.
\newblock {\em Journal of High Energy Physics}, 2011(8):1--31.

\bibitem[Fraser, 2011]{Fraser2011}
Fraser, D. (2011).
\newblock How to take particle physics seriously: A further defence of
  axiomatic quantum field theory.
\newblock {\em Studies In History and Philosophy of Modern Physics},
  42(2):126--135.

\bibitem[Gryb, 2012]{Gryb}
Gryb, S. (2012).
\newblock Shape dynamics and mach's principles: Gravity from conformal
  geometrodynamics.
\newblock {\em arXiv preprint arXiv:1204.0683}.

\bibitem[Haba, 2002]{BransDi2}
Haba, Z. (2002).
\newblock Renormalization in quantum brans-dicke gravity.
\newblock {\em arXiv preprint hep-th/0205130}.

\bibitem[Hagar, 2014]{Hagar}
Hagar, A. (2014).
\newblock {\em Discrete or continuous?: the quest for fundamental length in
  modern physics}.
\newblock Cambridge University Press.

\bibitem[Henson, 2009]{Henson}
Henson, J. (2009).
\newblock The causal set approach to quantum gravity.
\newblock {\em Approaches to Quantum Gravity: Towards a New Understanding of
  Space, Time and Matter}, 393.

\bibitem[Hossenfelder, 2013]{Hossenfelder}
Hossenfelder, S. (2013).
\newblock Minimal length scale scenarios for quantum gravity.
\newblock {\em Living Reviews in Relativity}, (2).

\bibitem[Jacobson, 1995]{Jacobson}
Jacobson, T. (1995).
\newblock Thermodynamics of spacetime: the einstein equation of state.
\newblock {\em Physical Review Letters}, 75(7):1260.

\bibitem[Jacobson and Parentani, 2003]{Jacobson2015}
Jacobson, T. and Parentani, R. (2003).
\newblock Horizon entropy.
\newblock {\em Foundations of Physics}, 33(2):323--348.

\bibitem[Kiefer, 2007]{KieferBook}
Kiefer, C. (2007).
\newblock Why quantum gravity?
\newblock In {\em Approaches to Fundamental Physics}, pages 123--130. Springer.

\bibitem[Liberati and Maccione, 2009]{Liberati}
Liberati, S. and Maccione, L. (2009).
\newblock Lorentz violation: motivation and new constraints.
\newblock {\em arXiv preprint arXiv:0906.0681}.

\bibitem[Malament, 1977]{Malament1977}
Malament, D.~B. (1977).
\newblock The class of continuous timelike curves determines the topology of
  spacetime.
\newblock {\em Journal of mathematical physics}, 18(7):1399--1404.

\bibitem[Niedermaier and Reuter, 2006]{NiedermaierReuter}
Niedermaier, M. and Reuter, M. (2006).
\newblock The asymptotic safety scenario in quantum gravity.
\newblock {\em Living Rev. Rel}, 9(5):173.

\bibitem[Orlando and Reffert, 2009]{Horava}
Orlando, D. and Reffert, S. (2009).
\newblock The renormalizability of hořava–lifshitz-type gravities.
\newblock {\em Classical and Quantum Gravity}, 26(15):155021.

\bibitem[Padmanabhan, 2011]{Padmanabhan2}
Padmanabhan, T. (2011).
\newblock Lessons from classical gravity about the quantum structure of
  spacetime.
\newblock In {\em Journal of Physics: Conference Series}, volume 306, page
  012001. IOP Publishing.

\bibitem[Percacci and Vacca, 2010]{Percacci2010}
Percacci, R. and Vacca, G.~P. (2010).
\newblock Asymptotic safety, emergence and minimal length.
\newblock (27):245026.

\bibitem[Peskin and Schroeder, 1995]{PeskinSchroeder}
Peskin, M.~E. and Schroeder, D.~V. (1995).
\newblock Quantum field theory.
\newblock {\em The Advanced Book Program, Perseus Books Reading,
  Massachusetts}.

\bibitem[Reid, 2001]{Reid}
Reid, D.~D. (2001).
\newblock Discrete quantum gravity and causal sets.
\newblock {\em Canadian Journal of Physics}, 79(1):1--16.

\bibitem[Rovelli, 2007]{RovelliCanonical}
Rovelli, C. (2007).
\newblock {\em Quantum gravity}.
\newblock Cambridge University Press.

\bibitem[Rovelli and Vidotto, 2014]{Rovellicovariant}
Rovelli, C. and Vidotto, F. (2014).
\newblock {\em Covariant Loop Quantum Gravity: An Elementary Introduction to
  Quantum Gravity and Spinfoam Theory}.
\newblock Cambridge University Press.

\bibitem[Schwartz, 2014]{Schwartz}
Schwartz, M.~D. (2014).
\newblock {\em Quantum field theory and the standard model}.
\newblock Cambridge University Press.

\bibitem[Stelle, 1977]{Stelle}
Stelle, K. (1977).
\newblock Renormalization of higher-derivative quantum gravity.
\newblock {\em Physical Review D}, 16(4):953.

\bibitem['t~Hooft and Veltman, 1974]{HooftVeltmann}
't~Hooft, G. and Veltman, M. (1974).
\newblock One-loop divergencies in the theory of gravitation.
\newblock In {\em Annales de l'IHP Physique th{\'e}orique}, volume~20, pages
  69--94.

\bibitem[Verlinde, 2011]{Verlinde}
Verlinde, E. (2011).
\newblock On the origin of gravity and the laws of newton.
\newblock {\em Journal of High Energy Physics}, 2011(4):1--27.

\bibitem[Visser, 2002]{Visser}
Visser, M. (2002).
\newblock Sakharov's induced gravity: a modern perspective.
\newblock {\em Modern Physics Letters A}, 17(15n17):977--991.

\bibitem[Wallace, 2006]{Wallace2006}
Wallace, D. (2006).
\newblock In defence of naivete: The conceptual status of lagrangian quantum
  field theory.
\newblock {\em Synthese}, 151(1):33.

\bibitem[Wallace, 2011]{Wallace2011}
Wallace, D. (2011).
\newblock Taking particle physics seriously: A critique of the algebraic
  approach to quantum field theory.
\newblock {\em Studies In History and Philosophy of Modern Physics},
  42(2):116--125.

\bibitem[Weinberg, 1979]{Weinberg}
Weinberg, S. (1979).
\newblock Ultraviolet divergencies in quantum theories of gravitation.
\newblock In Hawking, S. and Israel, W., editors, {\em General relativity, an
  Einstein Centenary survey}, pages 790--831. Cambridge University Press,
  Cambridge.

\bibitem[W\"{u}thrich, 2005]{Wuthrich2005}
W\"{u}thrich, C. (2005).
\newblock To quantize or not to quantize: Fact and folklore in quantum gravity.
\newblock {\em Philosophy of Science}, 72:777--788.

\bibitem[W{\"u}thrich, 2012]{Wuthrich}
W{\"u}thrich, C. (2012).
\newblock The structure of causal sets.
\newblock {\em Journal for general philosophy of science}, 43(2):223--241.

\bibitem[Zee, 2010]{Zee}
Zee, A. (2010).
\newblock {\em Quantum field theory in a nutshell}.
\newblock Princeton University Press.

\end{thebibliography}
\bibliographystyle{apalike}

\end{document}